\documentclass[%
 reprint,
superscriptaddress,
 amsmath,amssymb,
 aps,
]{revtex4-2}

\usepackage{graphicx}
\usepackage{dcolumn}
\usepackage{bm}
\usepackage{multirow}

\begin{document}


\title{First JSNS$^2$ measurement of the electron neutrino flux using the $^{12}C(\nu_{e},e^{-}) ^{12}N_{g.s.}$ reaction}

\author{T. Dodo}
\affiliation{Research Center for Neutrino Science, Tohoku University, 6-3 Azaaoba, Aramaki, Aoba-ku, Sendai, Miyagi 980-8578, Japan}
\affiliation{Advanced Science Research Center, JAEA, 2-4 Shirakata, Tokai-mura, Naka-gun, Ibaraki 319-1195, Japan}
\author{M. K. Cheoun}
\affiliation{Department of Physics and OMEG Institute, Soongsil University, 369 Sangdo-ro, Dongjak-gu, Seoul, 06978, Korea}
\author{J. H. Choi}
\affiliation{Laboratory for High Energy Physics, Dongshin University, 67, Dongshindae-gil, Naju-si, Jeollanam-do, 58245, Korea}
\author{J. Y. Choi}
\affiliation{Department of Fire Safety, Seoyeong University, 1 Seogang-ro, Buk-gu, Gwangju, 61268, Korea}
\author{J. Goh}
\affiliation{Department of Physics, Kyung Hee University, 26, Kyungheedae-ro, Dongdaemun-gu, Seoul 02447, Korea}
\author{M. Harada}
\affiliation{J-PARC Center, JAEA, 2-4 Shirakata, Tokai-mura, Naka-gun, Ibaraki 319-1195, Japan}
\author{S. Hasegawa}
\affiliation{Advanced Science Research Center, JAEA, 2-4 Shirakata, Tokai-mura, Naka-gun, Ibaraki 319-1195, Japan}
\affiliation{J-PARC Center, JAEA, 2-4 Shirakata, Tokai-mura, Naka-gun, Ibaraki 319-1195, Japan}
\author{W. Hwang}
\affiliation{Department of Physics, Kyung Hee University, 26, Kyungheedae-ro, Dongdaemun-gu, Seoul 02447, Korea}
\author{H. I. Jang}
\affiliation{Department of Fire Safety, Seoyeong University, 1 Seogang-ro, Buk-gu, Gwangju, 61268, Korea}
\author{J. S. Jang}
\affiliation{GIST College, Gwangju Institute of Science and Technology, 123 Cheomdangwagi-ro, Buk-gu, Gwangju, 61005, Korea}
\author{K. K. Joo}
\affiliation{Department of Physics, Chonnam National University, 77, Yongbong-ro, Buk-gu, Gwangju, 61186, Korea}
\author{D. E. Jung}
\affiliation{Department of Physics, Sungkyunkwan University, 2066, Seobu-ro, Jangan-gu, Suwon-si, Gyeonggi-do, 16419, Korea}
\author{S. K. Kang}
\affiliation{School of Liberal Arts, Seoul National University of Science and Technology, 232 Gongneung-ro, Nowon-gu, Seoul, 139-743, Korea}
\author{Y. Kasugai}
\affiliation{J-PARC Center, JAEA, 2-4 Shirakata, Tokai-mura, Naka-gun, Ibaraki 319-1195, Japan}
\author{T. Kawasaki}%
\affiliation{Department of Physics, Kitasato University, 1 Chome-15-1 Kitazato, Minami Ward, Sagamihara, Kanagawa, 252-0329, Japan}
\author{E. M. Kim}
\affiliation{Department of Physics, Chonnam National University, 77, Yongbong-ro, Buk-gu, Gwangju, 61186, Korea}
\author{E. J. Kim}
\affiliation{Division of Science Education, Jeonbuk National University, 567 Baekje-daero, Deokjin-gu, Jeonju-si, Jeollabuk-do, 54896, Korea}
\author{S. Y. Kim}
\affiliation{Department of Physics, Chonnam National University, 77, Yongbong-ro, Buk-gu, Gwangju, 61186, Korea}
\author{S. B. Kim}
\affiliation{School of Physics, Sun Yat-sen (Zhongshan) University, Haizhu District, Guangzhou, 510275, China}
\author{H. Kinoshita}
\affiliation{J-PARC Center, JAEA, 2-4 Shirakata, Tokai-mura, Naka-gun, Ibaraki 319-1195, Japan}
\author{T. Konno}
\affiliation{Department of Physics, Kitasato University, 1 Chome-15-1 Kitazato, Minami Ward, Sagamihara, Kanagawa, 252-0329, Japan}
\author{D. H. Lee}
\email{leedh@post.kek.jp}
\affiliation{High Energy Accelerator Research Organization (KEK), 1-1 Oho, Tsukuba, Ibaraki, 305-0801, Japan}
\author{C. Little}
\affiliation{University of Michigan, 500 S. State Street, Ann Arbor, Michigan 48109, USA}
\author{T. Maruyama}
\affiliation{High Energy Accelerator Research Organization (KEK), 1-1 Oho, Tsukuba, Ibaraki, 305-0801, Japan}
\author{E. Marzec}
\affiliation{University of Michigan, 500 S. State Street, Ann Arbor, Michigan 48109, USA}
\author{S. Masuda}
\affiliation{J-PARC Center, JAEA, 2-4 Shirakata, Tokai-mura, Naka-gun, Ibaraki 319-1195, Japan}
\author{S. Meigo}
\affiliation{J-PARC Center, JAEA, 2-4 Shirakata, Tokai-mura, Naka-gun, Ibaraki 319-1195, Japan}
\author{D. H. Moon}
\affiliation{Department of Physics, Chonnam National University, 77, Yongbong-ro, Buk-gu, Gwangju, 61186, Korea}
\author{T. Nakano}
\affiliation{Research Center for Nuclear Physics, Osaka University, 10-1 Mihogaoka, Ibaraki, Osaka, 567-0047, Japan}
\author{M. Niiyama}
\affiliation{Department of Physics, Kyoto Sangyo University, Motoyama, Kamigamo, Kita-Ku, Kyoto-City, 603-8555, Japan}
\author{K. Nishikawa}
\affiliation{High Energy Accelerator Research Organization (KEK), 1-1 Oho, Tsukuba, Ibaraki, 305-0801, Japan}
\author{M. Y. Pac}
\affiliation{Laboratory for High Energy Physics, Dongshin University, 67, Dongshindae-gil, Naju-si, Jeollanam-do, 58245, Korea}
\author{B. J. Park}
\affiliation{Department of Physics, Kyungpook National University, 80 Daehak-ro, Buk-gu, Daegu, 41566, Korea}
\author{H. W. Park}
\affiliation{Department of Physics, Chonnam National University, 77, Yongbong-ro, Buk-gu, Gwangju, 61186, Korea}
\author{J. B. Park}
\affiliation{Department of Physics and OMEG Institute, Soongsil University, 369 Sangdo-ro, Dongjak-gu, Seoul, 06978, Korea}
\author{J. S. Park}
\affiliation{Department of Physics, Chonnam National University, 77, Yongbong-ro, Buk-gu, Gwangju, 61186, Korea}
\author{J. S. Park}
\affiliation{Department of Physics, Kyungpook National University, 80 Daehak-ro, Buk-gu, Daegu, 41566, Korea}
\author{R. G. Park}
\affiliation{Department of Physics, Chonnam National University, 77, Yongbong-ro, Buk-gu, Gwangju, 61186, Korea}
\author{S. J. M. Peeters}
\affiliation{Department of Physics and Astronomy, University of Sussex, Falmer, Brighton, BN1 9RH, United Kingdom}
\author{J. W. Ryu}
\affiliation{Department of Physics, Kyungpook National University, 80 Daehak-ro, Buk-gu, Daegu, 41566, Korea}
\author{K. Sakai}
\affiliation{J-PARC Center, JAEA, 2-4 Shirakata, Tokai-mura, Naka-gun, Ibaraki 319-1195, Japan}
\author{S. Sakamoto}
\affiliation{J-PARC Center, JAEA, 2-4 Shirakata, Tokai-mura, Naka-gun, Ibaraki 319-1195, Japan}
\author{T. Shima}
\affiliation{Research Center for Nuclear Physics, Osaka University, 10-1 Mihogaoka, Ibaraki, Osaka, 567-0047, Japan}
\author{C. D. Shin}
\affiliation{High Energy Accelerator Research Organization (KEK), 1-1 Oho, Tsukuba, Ibaraki, 305-0801, Japan}
\author{J. Spitz}
\affiliation{University of Michigan, 500 S. State Street, Ann Arbor, Michigan 48109, USA}
\author{F. Suekane}
\affiliation{Research Center for Neutrino Science, Tohoku University, 6-3 Azaaoba, Aramaki, Aoba-ku, Sendai, Miyagi 980-8578, Japan}
\author{Y. Sugaya}
\affiliation{Research Center for Nuclear Physics, Osaka University, 10-1 Mihogaoka, Ibaraki, Osaka, 567-0047, Japan}
\author{K. Suzuya}
\affiliation{J-PARC Center, JAEA, 2-4 Shirakata, Tokai-mura, Naka-gun, Ibaraki 319-1195, Japan}
\author{Y. Yamaguchi}
\affiliation{J-PARC Center, JAEA, 2-4 Shirakata, Tokai-mura, Naka-gun, Ibaraki 319-1195, Japan}
\author{I. S. Yeo}
\affiliation{Laboratory for High Energy Physics, Dongshin University, 67, Dongshindae-gil, Naju-si, Jeollanam-do, 58245, Korea}
\author{I. Yu}
\affiliation{Department of Physics, Sungkyunkwan University, 2066, Seobu-ro, Jangan-gu, Suwon-si, Gyeonggi-do, 16419, Korea}

\collaboration{JSNS$^2$ Collaboration}



\date{\today}

\begin{abstract}
  JSNS$^2$ (J-PARC Sterile Neutrino Search at J-PARC Spallation Neutron Source)
  is an experiment searching for sterile neutrinos through the observation
  of $\bar{\nu}_{\mu} \rightarrow \bar{\nu}_e$ appearance oscillations,
  using neutrinos produced by muon decay-at-rest. A key aspect of the experiment involves accurately understanding the neutrino flux and the quantities of pions and muons, which are progenitors of (anti)neutrinos, given that their production rates have yet to be measured.
  We present the first electron-neutrino flux measurement using 
  $^{12}\mathrm{C}(\nu_{e},e^{-}) ^{12}\mathrm{N}_{g.s.}$ reaction in JSNS$^2$, yielding a flux of
  (6.7 $\pm$ 1.6 (stat.) $\pm$ 1.7 (syst.)) $\times$ 10$^{-9}$~cm$^{-2}$~proton$^{-1}$ at the JSNS$^2$ detector location, located at 24~meters distance from the mercury target. This flux measurement is consistent with predictions from simulations based on hadron models.
\end{abstract}

\maketitle


\section{\label{sec:intro} Introduction }

The JSNS$^2$ experiment was proposed in 2013~\cite{CITE:JSNS2proposal} to investigate  short-baseline neutrino oscillations,
a phenomenon suggested by the results of several previous experiments~\cite{CITE:LSND, CITE:BEST, CITE:MiniBooNE2018, CITE:REACTOR}.
The main search mode of the experiment is the appearance oscillation $\bar{\nu}_{\mu} \rightarrow \bar{\nu}_e$, as suggested by the LSND experiment~\cite{CITE:LSND}.
However, an important parameter, the flux of $\bar{\nu}_{\mu}$, 
cannot be predicted by current theories or simulations, as there have been no 
precise measurements of the relevant pion production rates. This uncertainty in the flux significantly impacts the sensitivity of the experiment and its ability to detect oscillations.
Therefore, the \textit{in-situ} measurement of the $^{12}C(\nu_{e},e^{-}) ^{12}N_{\text{g.s.}}$ reaction (``CNgs'')
is crucial for the indirect measurement of the $\bar{\nu}_{\mu}$ flux 
under this condition~\cite{CITE:JSNS2proposal, CITE:JSNS2TDR}. This is because the sources of both $\nu_{e}$ and $\bar{\nu}_{\mu}$ are the same: decay-at-rest plus-charged muons ($\mu$DAR). The cross section measurements from other experiments~\cite{CITE:KARMEN, CITE:LSND2} are used to relate  the $\nu_{e}$ and $\bar{\nu}_{\mu}$ fluxes.

The electron neutrino 
could also undergo oscillations, but the best-fit value for the mixing angle from reactor experiments~\cite{CITE:REACTOR} is relatively small ($\sim 10\%$). Therefore, in this manuscript, we assume no $\nu_{e}$ disappearance due to the limited precision of the measurement. 

The CNgs signal consists of two components: the prompt $e^{-}$ reaction and the delayed beta decay signal, $^{12}N \to e^{+} + \nu_{e} + ^{12}C$. The beta plus decay has the lifetime of 15.9~ms, and an end point of 16.3~MeV for the positron energy spectrum. This coincidence of prompt and delayed signals significantly reduces the accidental backgrounds, allowing JSNS$^2$ to efficiently detect the CNgs signal.

The JSNS$^2$ experiment uses the liquid scintillator detector located on the third floor of the Material and Life Science Experimental Facility (MLF) at J-PARC. The detector is positioned 24~meters from the mercury target.
It contains a Gadolinium-loaded liquid scintillator (Gd-LS) with a 0.1 \% Gd concentration, and 10\% by volume of di-isopropylnaphthalene (DIN,C$_{16}$H$_{20}$) is dissolved into the Gd-LS.
(Anti)neutrinos are produced at the mercury target through collisions with 1~MW, 3~GeV protons accelerated by a rapid cycling synchrotron with 25~Hz repetition 
in the MLF. The proton beam consists of two bunches with a 40~ns width (1 sigma of a Gaussian) and the bunches are separated by 600 ns each.

\section{\label{sec:detector} JSNS$^2$ detector}

The JSNS$^2$ detector is described in \cite{CITE:JSNS2NIM}. 
However, the components relevant for the CNgs
measurement are presented in this section.

The detector is cylindrical,
with a diameter of 4.6~m and height of 3.5~m, enclosed in a  stainless steel tank~\cite{CITE:HINOss}. Inside the tank, an acrylic vessel with dimensions of 3.2~m in diameter and 2.5~m in height holds
17~metric tons of a liquid scintillator cocktail. This mixture consists of 90\% Gd-LS and 10\% DIN by volume.

Surrounding the acrylic vessel, a gamma catcher and a cosmic ray veto region, situated between the acrylic vessel and stainless steel tank, contains 33~tonnes of Gd-unloaded liquid scintillator (LS). 
The liquid scintillator uses linear alkylbenzene (LAB) as the base solvent, with 3~g/L 2,5-diphenyloxazole (PPO) as the flour (the first wavelength shifter) and 15~mg/L 1,4-bis(2-methylstyryl) benzene (bis-MSB) as the second one.

The LS volume is divided into two layers: the catcher and the veto, separated
by an optical separator made of black acrylic boards. The gamma catcher region, located inside the optical barrier, consists of a $\sim$25~cm thick LS layer surrounding the acrylic vessel.

Ninety-six Hamamatsu R7081 photomultiplier tubes (PMTs), each with a 10-inch diameter, detect light from the liquid scintillator in the acrylic vessel and catcher regions. 
The veto layer, situated outside the optical barrier, is designed to detect particles entering the detector from outside the stainless steel tank. 

This veto later is equipped with twenty-four 10-inch PMTs, and reflective sheets cover the whole surface of the veto layer to enhance the collection efficiency of the scintillation light.

The signals from 120~PMTs are digitized and recorded at a 500~MHz sampling rate using 8-bit flash analog-to-digital converters (FADCs)~\cite{CITE:JSNS2DAQ}. The JSNS$^{2}$ experiment implements a sequence and coincidence trigger to detect inverse-beta-decay (IBD) from  $\bar{\nu}_{\mu} \rightarrow \bar{\nu}_e$ 
oscillation and CNgs signals. 

A trigger, referred to as the ``prompt trigger," is generated if an analog sum of the 96~PMTs in the acrylic and catcher regions exceeds a 
threshold of approximately 200~mV ($\sim 5$~MeV) within a window of 1.7--10~$\mu s$ relative to the beam timing.

After the prompt trigger is activated, a delayed trigger is generated when the analog sum exceeds a different threshold of approximately 70~mV  ($\sim 2$~MeV) within a time
window of 12~ms (2021 data) to 25~ms (2022 data)  after the prompt trigger.  

The beam timing is synchronized with the accelerator's scheduled timing via a radio-frequency (RF) module, which operates on a 40~ms cycle. 

\section{Dataset}

Following the commissioning run in 2020~\cite{CITE:EPJC}, JSNS$^2$ has conducted 
stable data taking from 2021 to 2024. During this period, the beam power increased from 0.63~MW to 0.95~MW, resulting in a total accumulated proton-on-target (POT) of 4.85 $\times$ 10$^{22}$. 

For this analysis, data corresponding to $2.2 \times 10^{22}$~POT are used, with $1.29\times 10^{22}$ in 2021 and $0.91\times 10^{22}$ in 2022.

\section{Event Properties}

The theoretical cross section and properties of the CNgs reactions are described in~\cite{CITE:Fukugida}, while the cross sections have been measured by other 
experiments~\cite{CITE:KARMEN, CITE:LSND2}.   

\begin{figure}[htbp]
	\centering
	\begin{minipage}[b]{0.395\textwidth}
      \includegraphics[width=0.924\textwidth]{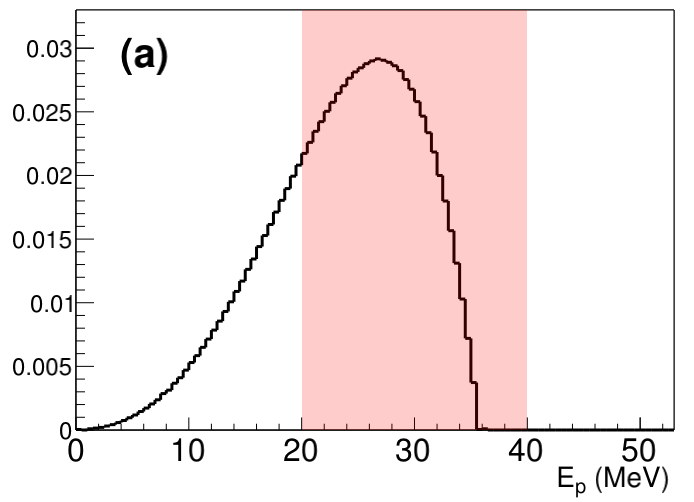}
			\qquad
	\end{minipage}
	\begin{minipage}[b]{0.392\textwidth}
       \includegraphics[width=0.924\textwidth]{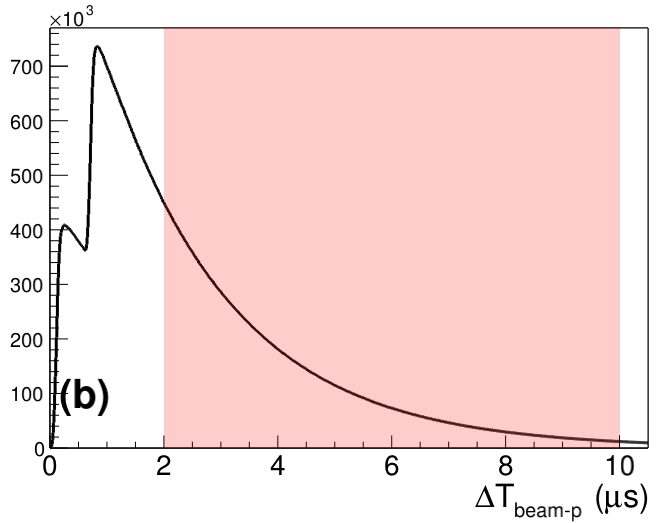}
			\qquad
    \end{minipage}
	\begin{minipage}[b]{0.392\textwidth}
       \includegraphics[width=0.924\textwidth]{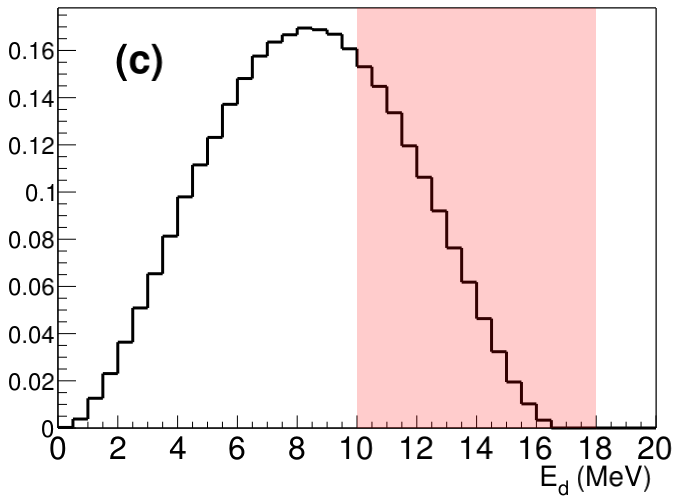}
			\qquad
	\end{minipage}        
    \begin{minipage}[b]{0.392\textwidth}
        \includegraphics[width=0.924\textwidth]{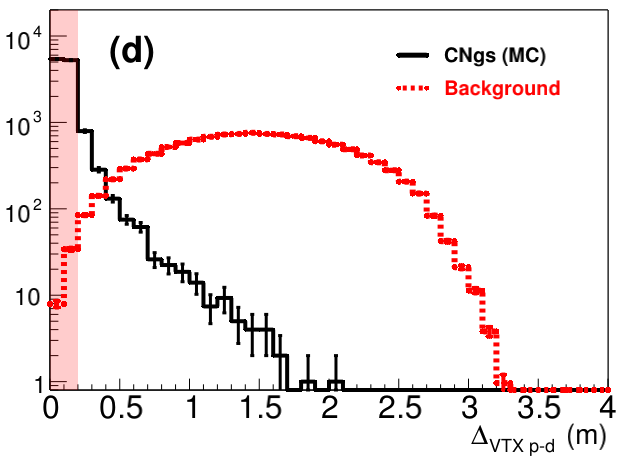}
			\qquad
	\end{minipage}
	\caption{Properties of CNgs events. (a) The energy of the prompt signal (MeV), 
(b) The timing of the prompt signal from the beam starting time ($\mu$s), (c) The energy of the delayed signal (MeV) and (d) The spatial correlation between the prompt and delayed signals (meter). The vertical axes of all plots are in arbitrary units. The shaded regions are used for the event selection that is described later. The plot for the most important variable (d) to reject the background includes its distribution here. Background distributions of (a)-(c) are relatively flat as shown later.}
\label{fig:properties}
\end{figure}

The energy spectrum of the $\nu_{e}$ from $\mu$DAR 
($d\Gamma/dE_{\nu_{e}})$ is expressed as
\begin{equation}
    \frac{d\Gamma}{dE_{\nu_{e}}} = \frac{G_F^2 m_{\mu}^2}{2 \pi^3} E_{\nu_{e}}^2 \left(1-2 \frac{ E_{\nu_{e}}}{m_{\mu}}\right) \hspace{0.5truecm}\left(E_{\nu_{e}}\le\frac{m_{\mu}}{2}\right)
\end{equation}
, where $E_{\nu_{e}}$ is the energy of $\nu_{e}$, $m_{\mu}$ is the invariant mass of the muon and $G_{F}$ is Fermi coupling constant.
The $Q$-value of the reaction is 17.3~MeV, and the differential cross section for the
CNgs reaction is provided in~\cite{CITE:Fukugida}. The energy spectrum of the prompt electron signal based on theoretical predictions is shown in Fig.~\ref{fig:properties}(a).

The timing of the prompt signal relative to the beam start time, as obtained from simulation, is shown in Fig.~\ref{fig:properties}(b).
The proton beam striking the mercury target consists of two bunches separated by 600~ns. The simulation accounts for the muon lifetime at rest ($\sim$ 2.2 $\mu s$) . This timing selection is optimized to capture events from $\mu$DAR while effectively rejecting background signals from beam neutrons and other neutrinos.

The energy spectrum of the delayed positron signals from $^{12}N_{g.s.}$ beta-decay ($d\Gamma/dE_{e^{+}}$)
is given by:
\begin{gather}
\label{Eq:2}
    \frac{d\Gamma}{dE_{e^{+}}} = P_{e^{+}}E_{e^{+}} \left(E_{max} -E_{e^{+}}\right)^2\frac{2\pi\eta}{e^{2\pi\eta}-1},  \\
    (\eta=Z\alpha/\beta_{e^{+}}) \nonumber 
\end{gather}
Here, $E_{e^{+}}$ is the total positron energy, 
including rest energy, and $E_{max}$ is 16.83~MeV, which corresponds to the maximum energy of 
$^{12}N_{g.s.}$ beta-decay. The term, $2\pi\eta / (e^{2\pi\eta}-1)$, is the Fermi function to correct the Column effects.
$\eta$ is Sommerfeld parameter, where $Z$ is the particle's charge and $\alpha$ is the fine-structure constant. $\beta_{e^{+}}$ is the relativistic velocity, which is defined by $\beta_{e^{+}} = v_{e^{+}}/c$. 

Figure~\ref{fig:properties}(c) shows the
visible energy spectrum of the delayed positron ($E_{d}$), accounting for the annihilation energy within the detector in addition to Eq.~(\ref{Eq:2}). The mean lifetime of the beta decay is 15.9~ms and this will also be incorporated
into the event selection later.

\section{Event selection}

The prompt and delayed triggered events are paired during the offline selection.
The selection criteria for CNgs based on the event properties described above, are 
summarized in Table~\ref{tab:EventSelection}. This table also includes the efficiencies and 
systematic uncertainties associated with each selection. 

Since the run conditions, including triggers, differed between 2021 and 2022, the efficiencies of CNgs are listed separately for each year. The efficiency and systematic uncertainties are estimated using Monte Carlo (MC) simulations. 

The shaded regions in Fig.~\ref{fig:properties} indicate the event selection regions.

The trigger efficiency for prompt signals with an energy larger than 20~MeV is approximately 100\%, defining the energy threshold for the prompt signal. JSNS$^2$ uses the
Gd-LS, which emits gamma rays with a total energy of approximately 8~MeV when thermal neutrons generated by cosmogenic sources are captured by Gd. Consequently, the energy threshold for the delayed signal is set at 10~MeV.

The in-situ energy calibration has been performed using cosmogenic sources, 
including neutron capture events on gadolinium, which produce gamma rays with a total energy of approximately 8~MeV, and the end point of the Michel electrons at around 53~MeV. These calibrations account for time variations 
(with energy scale corrections below 2\%) and the position-dependent effects in the detector 
(variations less than 4\%) based on these reference samples. Additionally, systematic uncertainties account for nonlinearity of light yield caused by scintillation quenching 
between 8 to 53~MeV. Details of these calibrations are provided in Ref.~\cite{CITE:Cf}.

The timing criteria for prompt signals relative to the beam start, as well as the prompt-to-delayed interval ($\Delta T_{p-d}$), are determined by the trigger conditions.  
The FADC timing is calibrated using a 25~Hz beam, ensuring that time-related uncertainties  are negligible compared to those arising from energy considerations.

Beyond energy and timing selections, a spatial correlation between prompt and delayed signals is imposed ($\Delta_{VTX} \le$ 20 cm), based on reconstructed vertex positions. The event reconstruction algorithm used in JSNS$^2$ is detailed in references~\cite{CITE:Johnathon, CITE:Cf}.
This $\Delta_{VTX}$ cut is the most critical 
selection criterion. Figure~\ref{fig:properties}(d) shows the area-normalized $\Delta_{VTX}$ distribution for CNgs signal and background events. 
Due to the phase space, the $\Delta_{VTX}$ cut achieves a background rejection power of approximately 300 while maintaining a CNgs signal efficiency of about 90\%. 

To reject cosmic muons, events are required to have a sum of charges from the top 12 and bottom 12 veto PMTs below 30~p.e. and 40~p.e., respectively. Some background for the prompt signal arises from Michel electrons; therefore,  parent muons are searched for within 10 $\mu$s prior to the prompt candidate. Parent muons
are identified as events with more than 100 p.e. in either the top or 
bottom 12 veto PMTs. These criteria provide good rejections of cosmic 
rays better than 99\%, however also gives the some efficiency losses 
of CNgs signal due to the accidental coincidence between muons and the CNgs signal as shown in Table~\ref{tab:EventSelection}. 

The fiducial volume is defined by $R < 140$~cm and $|z| < 100$~cm, where the origin of the coordinate system is the detector center,
and $R = \sqrt{x^2 + y^2}$. This selection minimizes external background contamination.

\begin{table}[htb]
    \centering
    \caption{\label{tab:EventSelection}
    The CNgs selection criteria and their efficiencies. The efficiencies of CNgs signals are estimated using a simulation. }
    \vspace{3pt}
    \begin{tabular}{ccc}\hline
        Requirement & Efficiency (2021) & (2022) \\\hline
        --Prompt Event--\\

        $20\le E_{\mathrm{p}} \le 40$~MeV     & 67.3$\pm$1.9\% & 67.3$\pm$2.2\%\\
        $2.0\le \Delta T_{beam-p}\le 10$~$\mu$s & 49.8$\pm$0.2 & 46.8$\pm$0.1 \\
                                                            & &    \\

        --Delayed Event-- \\
        $10\le E_{\mathrm{d}}\le 18$~MeV      & 37.9$\pm$1.4 & 37.9$\pm$1.4 \\
                                                            &    & \\

        --CNgs paired Event-- \\
        0.2 $\le \Delta T_{p-d} \le 12 ~or ~25 $~ms & 51.8$\pm$0.04 & 78.1$\pm$0.04\\
        $\Delta_{VTX ~ p-d} \le 20$~cm               & 88.1$\pm$0.7 & 88.1$\pm$0.7 \\ 
                                                            &    &  \\
                                                            
         --Muon rejections-- \\
        Muon rejection                       & 88.5$\pm$0.4 & 90.1$\pm$0.3 \\
        Michel electron rejections           & 97.0$\pm$0.03 & 97.0$\pm$0.03\\ \hline 
        
        Total                                & 4.97$\pm$0.24 & 7.17$\pm$0.37\\\hline
    \end{tabular}
\end{table}
With these selection criteria, a total of 79 events were selected in the datasets from 2021 and 2022. The average event selection efficiency, normalized by POT, is 5.88$\pm$0.21\%.

\section{Background estimation}

A data-driven method is implemented to estimate the background. Fake trigger structures are created within the "delayed trigger" time 
window, which is set to 12~ms for the 2021 dataset and 25~ms for the 2022 dataset. Since the CNgs prompt signal occurs within the $\mu$DAR timing region, the fake prompt trigger timing is chosen to start at least 1~ms from the beam to avoid the contamination from actual CNgs events. 

To improve statistical precision, the fake beam start time is shifted by 10~$\mu$s increments  for each estimation. Using this method, the shape of the $\Delta_{VTX}$ distribution for the "background" in Fig.~\ref{fig:properties}(d) is derived. The number of background events within the CNgs selection region is estimated by the normalizing the $\Delta_{VTX}$ distributions for separations greater than 40~cm.

\begin{figure}
\includegraphics[width=0.5\textwidth]{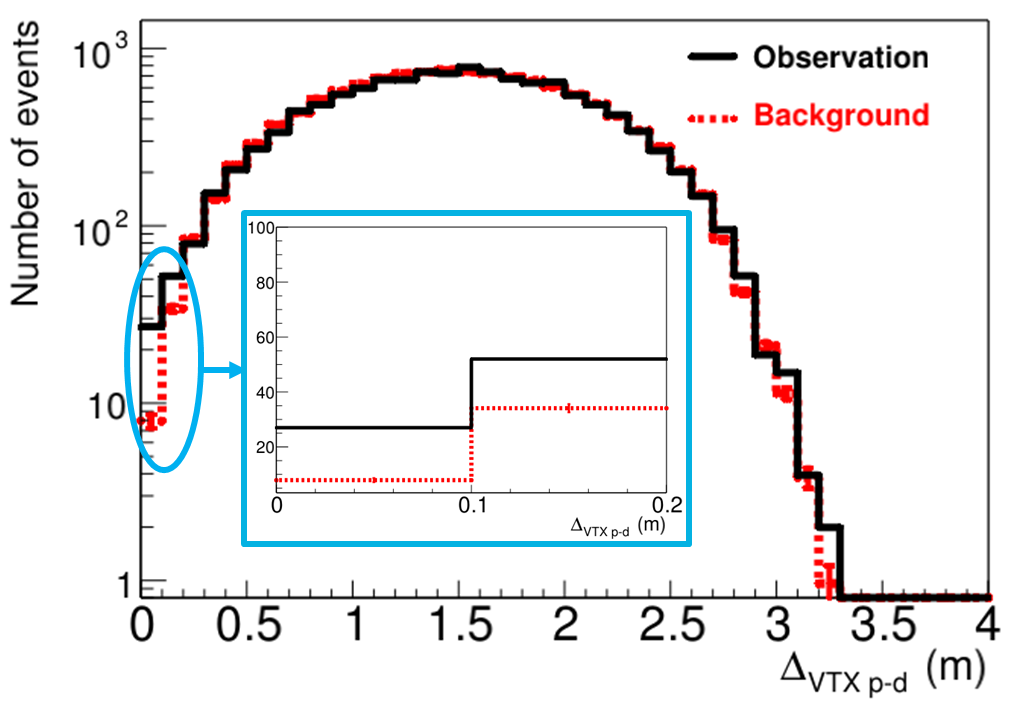} 
\caption{\label{fig:DVTX} The $\Delta_{VTX}$ distributions for the selected events (black) and the estimated background using the fake trigger method (red dashed) are shown.  The vertical axis represents the number of events. All event selections, except for the $\Delta_{VTX}$ cut, are applied. An inset linear scale plot highlights the region where $\Delta_{VTX} < 20$~cm.
} 
\end{figure}

Figure~\ref{fig:DVTX} presents the $\Delta_{VTX}$ distribution
after normalization using events with $\Delta_{VTX} > 40$~cm. All other selection criteria, except for the $\Delta_{VTX}$ cut, are applied in this plot. A clear excess of events is seen in the CNgs selection region.
This method provides in-situ background estimations, making it free from the systematic uncertainties related to energy, timing, 
and muon rejection criteria.

As an alternative method to estimate the accidental background events, the spill-shift method is employed. This method uses a different 40~ms time window initiated by the beam timing.
Once a CNgs prompt candidate is identified in the prompt trigger, delayed activities are searched within a different 40~ms beam spill time span. Details of this method are described in reference~\cite{CITE:accidental}, which estimates the accidental background for the CNgs events. 

Table~\ref{tab:Comp_data_MST_n} summarizes the selected number of candidates and the estimated background events. 

\begin{table}[h]
    \caption
    {
        A comparison between the number of selected events and the number of estimated background events.
    }
    \vspace{0.2truecm}
    \label{tab:Comp_data_MST_n}
    \centering
    \begin{tabular}{cccc}
        \hline
         & observation & fake method & spill shift\\
        \hline
        $\Delta_{VTX}$~$< 20$~cm & 79 & $42.2\pm1.7$ & $37.7\pm0.6$ \\
        \hline
    \end{tabular}
\end{table}

A 10.8\% difference is observed between the two background estimation methods. The fake trigger method is considered to provide a more accurate estimation, as it may better account for cosmogenic correlated background.
Consequently, the number derived from the fake trigger method is used as the central value,
and the observed difference is treated as a systematic uncertainty in the background estimation. The total number of background events is determined to be $42.2\pm4.8$. 

The $p$-value, calculated based on this background estimation, is 2.9$\times$10$^{-7}$.
The result rejects the hypothesis that the observed number of events is due to background fluctuations with a significance exceeding 6~$\sigma$.

\section{Comparison between observation and prediction}

\begin{figure}[htbp]
	\centering
	\begin{minipage}[b]{0.41\textwidth}
        \includegraphics[width=0.93\textwidth]{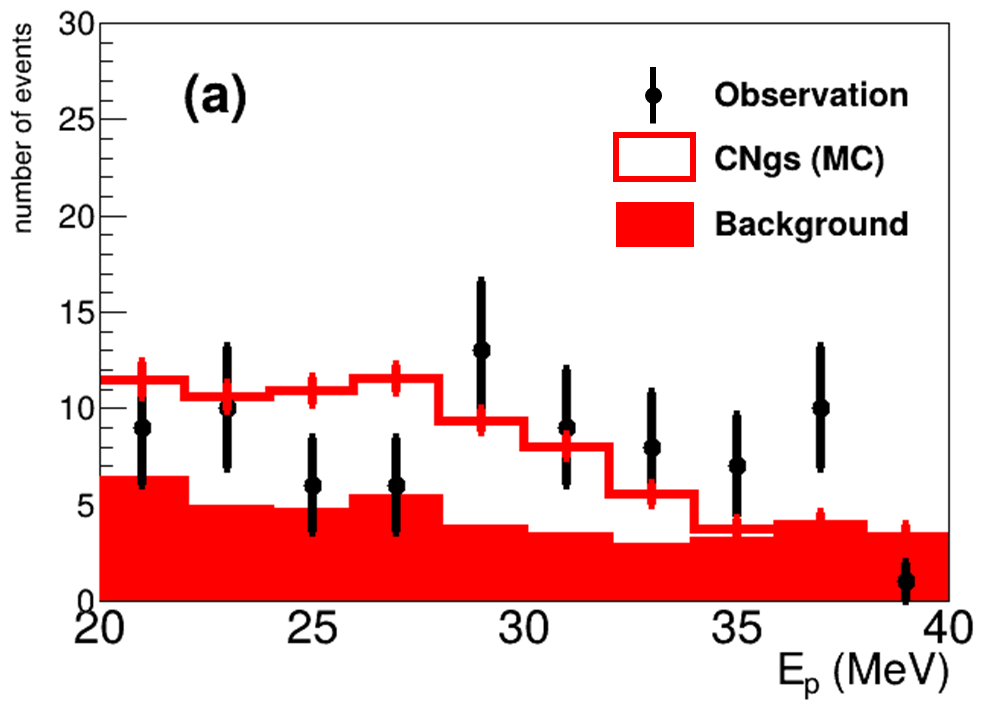}
			\qquad
	\end{minipage}
	\begin{minipage}[b]{0.41\textwidth}
       \includegraphics[width=0.99\textwidth]{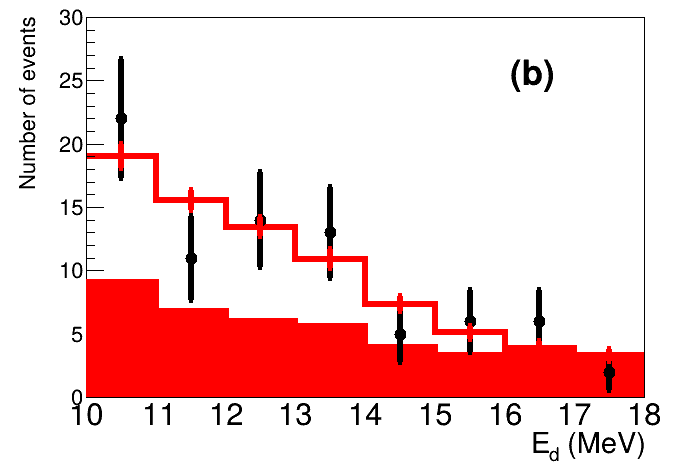}
			\qquad
    \end{minipage}
	\begin{minipage}[b]{0.41\textwidth}
       \includegraphics[width=0.93\textwidth]{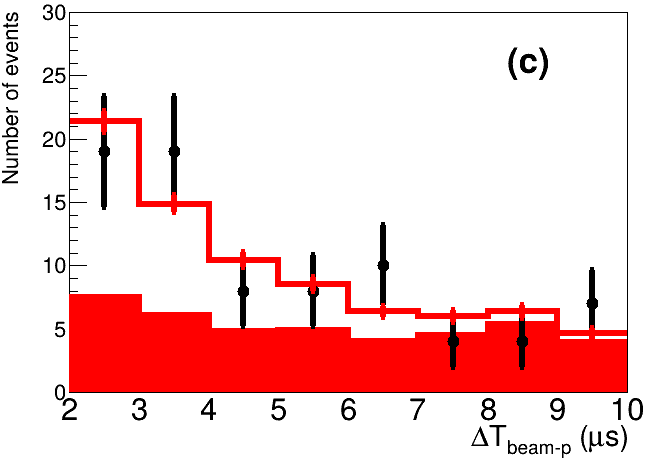}
			\qquad
	\end{minipage}        
    \begin{minipage}[b]{0.41\textwidth}
        \includegraphics[width=0.985\textwidth]{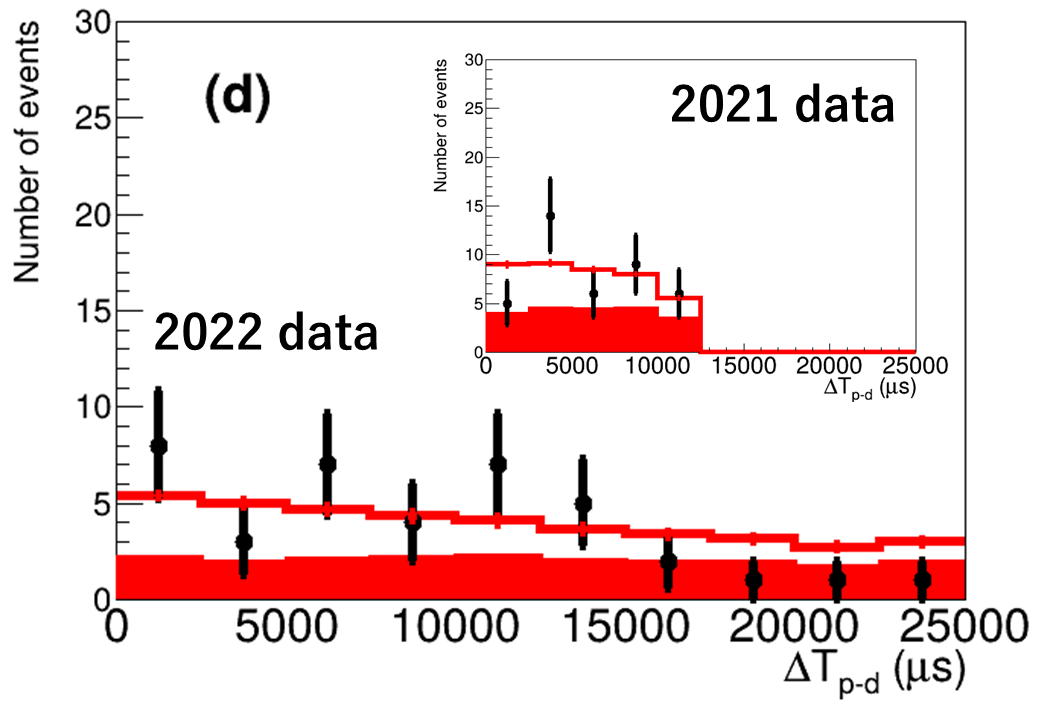}
			\qquad
	\end{minipage}
	\caption{A comparison is shown between the selected events (black points with error bars) 
and the predictions (histograms) for energy and timing distributions. 
The vertical axis represents the number of events. 
Panel (a) shows the energy distribution of prompt candidates,
(b) the energy of delayed candidates, (c) the timing of prompt candidates relative to the beam, and (d) the $\Delta T_{p-d}$ distribution. Due to the differences in the delayed candidate trigger windows between 2021 and 2022,  separated plots are provided for each dataset.}
\label{fig:obs_vs_pred}
\end{figure}

Figure~\ref{fig:obs_vs_pred} compares the  selected events (black points with error bars) to the predictions for energy and timing distributions (histograms). The prediction combines the CNgs signal from Monte Carlo simulations (white) and the data-driven background estimation (red). The signal-to-background ratio, based on
Table~\ref{tab:Comp_data_MST_n}, is 36.8 signal events and the 46.6 background events. The background component is highlighted in red shading.

The predictions and observations show good agreement.

\section{Neutrino flux determination}

The electron neutrino flux ($\Phi$) is given by the following formula:
\begin{equation}
    \Phi = \frac{N_{\mathrm{eve}}}{\sigma_{CNgs} \cdot N_{\mathrm{target}} \cdot \epsilon}, 
\label{eq:flux}
\end{equation}
where $N_{\mathrm{eve}}$ represents the number of observed signal events, $\sigma_{CNgs}$ is the cross section for the CNgs reaction, $N_{\mathrm{target}}$ is the number of carbon targets ($^{12}$C) within the fiducial volume,
and $\epsilon$ is the selection efficiency for CNgs events. The value $\epsilon$ and its associated uncertainties have been discussed earlier. Further details for each parameter are provided in this section.

\subsection{The number of signal events ($N_{\mathrm{eve}}$)}

The number of CNgs signal events is determined using the values in Table~\ref{tab:Comp_data_MST_n}. A statistical uncertainty is calculated as $\sqrt{79} = 8.9$, resulting in a total number of CNgs event count of $36.8\pm8.9$~(stat.)~$\pm4.8$~(syst.). The systematic uncertainty is primarily due to the difference of two background estimation methods. 

\subsection{The cross section ($\sigma_{CNgs}$)}

The weighted average of the measured cross sections, $(9.1\pm0.7)\times 10^{-42}$~cm$^{2}$, is calculated using values from 
KARMEN~\cite{CITE:KARMEN}: $(9.3\pm0.9)\times 10^{-42}$~cm$^{2}$ and LSND~\cite{CITE:LSND2}: $(8.9\pm0.9)\times 10^{-42}$~cm$^{2}$. 

\subsection{The number of $^{12}$C targets ($N_{\mathrm{target}}$)}

The scintillation cocktail in the JSNS$^2$ target consists of 90\% (by volume) 
LAB-based Gd-LS and 10\% DIN
as previously mentioned. The fiducial volume is defined by the conditions $|z| < 100$~cm and $R < 140$~cm, which corresponds to a total volume of $1.23 \times 10^{4}$~L. The densities of  LAB and DIN are 0.860~g/mL~\cite{CITE:DB} and 
0.958 g/mL~\cite{CITE:EJ}, respectively. 
The weight ratio of the chemical components in Gd-LS is C:H:Gd = 87.7:12.1:0.103\%~\cite{CITE:DB}. Note that the Gd-LS used in JSNS$^2$ was donated by the Daya-Bay experiment. The chemical formula of DIN is C$_{16}$H$_{20}$, and the natural abundance of the carbon is 99:1 for $^{12}$C and $^{13}$C.

Based on those parameters, the number of $^{12}$C atoms in the fiducial volume is calculated to be $4.68\times 10^{29}$. The uncertainty in the number of target is primarily driven by the uncertainty in the fiducial volume, which results from the vertex reconstruction capability. Reference~\cite{CITE:Cf} discusses
the fiducial volume's uncertainties, which are estimated to be 20\%. Therefore, the number of $^{12}$C target is $(4.68\pm 0.94) \times10^{29}$ with this uncertainty.

\section{Neutrino flux ($\Phi$)}

\begin{figure}
\includegraphics[width=0.48\textwidth]{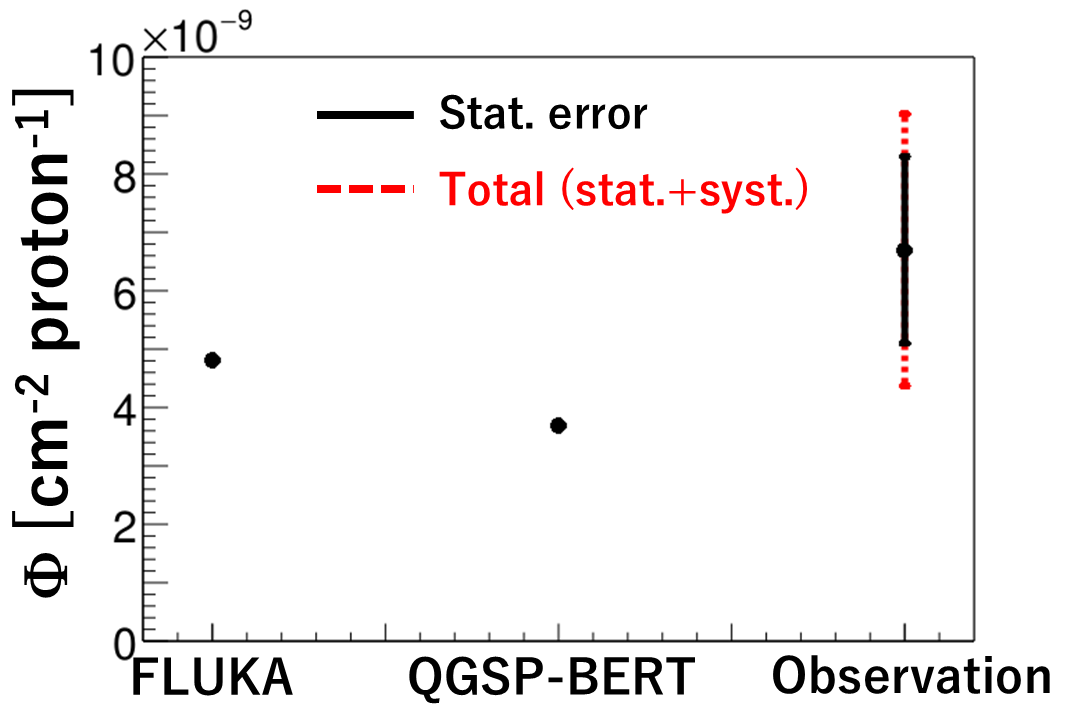} 
\caption{\label{fig:flux} 
A comparison between the observed electron neutrino flux and the flux predicted by simulation at the JSNS$^2$ detector location. The black bar for the observation corresponds to the statistical error and the red dashed one shows the total uncertainty (stat.+syst.).
} 
\end{figure}

Using Eq.(\ref{eq:flux}) with provided parameters, the electron neutrino flux at the JSNS$^2$ detector location, 24~m from the mercury target, is determined to be 
$[6.7\pm 1.6$~(stat.)$\pm~1.7$~(syst.)$] \times 10^{-9}$~cm$^{-2}$~proton$^{-1}$.
This result is consistent with the predictions in the JSNS$^2$ proposal~\cite{CITE:JSNS2proposal} and the
TDR~\cite{CITE:JSNS2TDR}, which uses FLUKA~\cite{CITE:FLUKA1,CITE:FLUKA2} and GEANT4 (QGSP-BERT)~\cite{CITE:Geant4} simulations.
FLUKA predicts $4.8\times 10^{-9}$~cm$^{-2}$~proton$^{-1}$ and QGSP-BERT predicts $3.7\times 10^{-9}$~cm$^{-2}$~proton$^{-1}$
 (Fig.~\ref{fig:flux}).
The measured values are 1.4 to 1.8~times larger than the simulation predictions, with
a significance ranging from 0.82 to 1.3 sigma.  

\section{\label{sec:conclusion} Conclusion}

JSNS$^2$ has conducted the first measurement of the electron neutrino flux via the $^{12}C(\nu_{e},e^{-}) ^{12}N_{g.s.}$
reaction, which provides a robust double-coincidence signal. The measured $\nu_{e}$ flux,  
$[6.7\pm1.6$~(stat.)~$\pm 1.7$~(syst.)$]\times 10^{-9}$~cm$^{-2}$~proton$^{-1}$, is consistent with simulation predictions within a significance range of 0.82 to 1.3 sigma.  

The number of electron neutrinos at the target, corresponding to the number of parent muons
($\mu^{+}$), is $0.48\pm0.17$ per proton with multiplying the 4$\pi r^2 (r = $2400 cm, which corresponds to the distance between the detector and the mercury target) . This value is approximately 10 times larger than that of KARMEN~\cite{CITE:KARMEN2}, attributed to the higher energy of the proton beam. This measurement also serves as a basis for 
the $\bar{\nu}_{\mu} \rightarrow \bar{\nu}_e$ search, representing the $\bar{\nu}_{\mu}$ flux.

Updated results with improved statistics, utilizing the current and the future POTs, and with the improved systematic uncertainty on the fiducial volume are highly desirable. 
In the future, JSNS$^2$-II~\cite{CITE:JSNS2-II}, which uses two detectors, will prove further clarification of the produced electron neutrino flux. Additionally, a hadron production rate measurement using a 3-GeV proton beam and a mercury target is recommended to further improve these results.

\hspace{0.7cm}

\begin{acknowledgments}
We deeply thank the J-PARC for their continuous support, 
especially for the MLF and the accelerator groups to provide 
an excellent environment for this experiment.
We acknowledge the support of the Ministry of Education, Culture, Sports, Science, and Technology (MEXT) and the JSPS grants-in-aid: No.\,16H06344, No.\,16H03967, No.\,23K13133, 
No.\,24K17074 and No.\,20H05624, Japan. This work is also supported by the National Research Foundation of Korea (NRF): No.\,2016R1A5A1004684, No.\,17K1A3A7A09015973, No.\,017K1A3A7A09016426, No.\,2019R1A2C3004955, No.\,2016R1D1A3B02010606, No.\,017R1A2B4011200, No.\,2018R1D1A1B07050425, No.\,2020K1A3A7A09080133, No.\,020K1A3A7A09080114, No.\,2020R1I1A3066835, 
No.\,2021R1A2C1013661, 
No.\,NRF-2021R1C1C2003615, No.\,2021R1A6A1A03043957, No.\,2022R1A5A1030700, 
No.\,RS-2023-00212787, and No.\,RS-2024-00416839. Our work has also been supported by a fund from the BK21 of the NRF. The University of Michigan gratefully acknowledges the support of the Heising-Simons Foundation. This work conducted at Brookhaven National Laboratory was supported by the U.S. Department of Energy under Contract DE-AC02-98CH10886. The work of the University of Sussex is supported by the Royal Society grant No.\,IESnR3n170385. We also thank the Daya Bay Collaboration for providing the Gd-LS, the RENO collaboration for providing the LS and PMTs, CIEMAT for providing the splitters, Drexel University for providing the FEE circuits and Tokyo Inst. Tech for providing FADC boards.
\end{acknowledgments}

\hspace{0.7cm}

\section*{DATA AVAILABILITY}
The data that support the findings of this article are not publicly available. The data are available from the authors upon reasonable request.

\nocite{*}

\bibliography{apssamp}

\end{document}